\documentclass[prb,twocolumn,showpacs,preprintnumbers,amsmath,amssymb]{revtex4}
\usepackage[dvips]{graphics}
\usepackage{epsfig}

\begin{document}

\title{Proximity of LaOFeAs to a magnetic instability}

\author{I. Opahle}
\author{H. C. Kandpal}
\author{Y. Zhang}
\author{C. Gros}
\author{R. Valent\'{i}}

\affiliation{
Institut f\"ur Theoretische Physik, J. W. Goethe-Universit\"at Frankfurt, \\
Max-von-Laue Strasse 1, 60438 Frankfurt/Main Germany}

\date{\today}

\begin{abstract}
We investigate the effect of external pressure 
on the Fe magnetic moment in undoped LaOFeAs within the framework of density functional theory 
and show that this system is close to a magnetic instability:
The Fe moment is found to drop by nearly a factor of 3 within a pressure range of $\pm$ 5 GPa
around the calculated equilibrium volume.
While the Fe moments show an unusually strong sensitivity to the spin arrangement (type of anti-ferromagnetic structure), the low temperature structural distortion is found to have only a minor influence on them.
Analysis of the Fermi surface topology and nesting features shows that these properties change
very little up to pressures of at least 10 GPa.
We discuss the magnetic instability in terms of the itinerancy of this system.
\end{abstract}

%\keywords{density functional theory, LaOFeAs, pressure}
\maketitle

%%%%%%%%%%%%%%%%%%%%%%%% INTRODUCTION %%%%%%%%%%%%%%%%%%%%%%%%%%%%%%%
\section{Introduction}

The discovery of high $T_c$ superconductivity in fluorine doped LaOFeAs~\cite{Kam08} 
with a critical temperature $T_c$ of about 26 K has stimulated an
enormous interest in these compounds. Shortly after this discovery it became clear that a whole
family of related compounds shows superconductivity at elevated temperatures. 
Substitution of La by other rare earth elements increases $T_c$ up to about 50~K~\cite{Ren08} and superconductivity at 38 K was also observed in the related K$_{0.4}$Ba$_{0.6}$Fe$_2$As$_2$ compound~\cite{Rot08}.

The undoped parent compound LaOFeAs is an anti-ferromagnet with a small ordered Fe moment of about 0.4~$\mu_B$~\cite{Cru08}. Density functional theory (DFT) calculations find, in contrast, a much too large value for the Fe moment, close to 2~$\mu_B$. The failure of DFT calculations to describe the Fe magnetic moment in these systems has raised doubts whether the underlying electronic structure is correct and whether it provides a sound basis for discussion of the superconducting state in the doped compounds. We argue here that the discrepancies between experiment and theory have a physical origin, namely the fact that LaOFeAs is close to a magnetic instability. Based on our DFT calculations we show that the Fe moment is highly susceptible to external pressure and drops by almost a factor of 3 within the pressure range from -5 to 5 GPa. This drastic change of the Fe moment goes along with only subtle changes in the electronic structure, what explains the initially apparent differences between DFT calculated moments and experimental observations. The predicted changes of the Fe moment allow for direct experimental verification, either by applying hydrostatic pressure or negative pressure which could be realized by hydrogenation.

\begin{figure}
\includegraphics[width=0.43\textwidth]{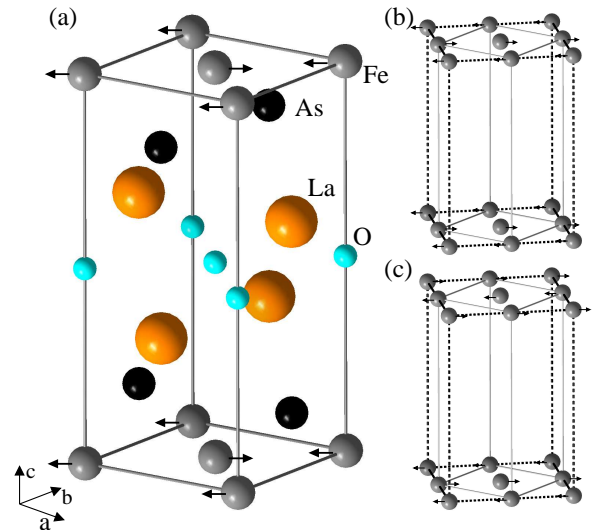}
\caption{\label{FIG:af123}(Color online)
Crystal structure of LaOFeAs (a) and spin arrangement of the three anti-ferromagnetic 
structures AF1 (a), AF2 (b) and AF3 (c)
considered in the calculations.
The unit cell for AF3 is doubled along the $c$-direction.}
\end{figure}

The crystal structure of LaOFeAs is tetragonal at room temperature and consists of FeAs layers separated by LaO layers (Fig.~\ref{FIG:af123}). Below $T_s\approx$ 155 K a weak structural distortion is observed, followed by the formation of a spin density wave (SDW) state around $T_N\approx$ 137 K~\cite{Cru08}. The low temperature crystal structure has been described as either monoclinic~\cite{Cru08} or orthorhombic~\cite{Nom08}. Both structures differ only marginally from each other, so that the symmetry can be described as orthorhombic (see supplement of Ref.~\onlinecite{Cru08}).
The anti-ferromagnetic (AF) order of most of the undoped parent compounds of the iron arsenide superconductors
has lead to speculations that spin fluctuations could be decisive for the pairing 
mechanism~\cite{Wang08}. On the other hand, the absence of a SDW state has been reported
for NdFeAsO, where AF order is only observed below 2 K and Fe orders together with the
Nd moments~\cite{Qiu08}.
A strong electron-phonon coupling of the Fe breathing mode in LaO$_{1-x}$F$_{x}$FeAs
was reported in Ref.~\onlinecite{Esc08}, which could contribute to the 
high $T_c$.
Thus, the pairing mechanism is still under debate and requires
further experimental and theoretical studies~\cite{Cve08}.

The effect of hydrostatic pressure on the magnetic and superconducting 
properties of the iron arsenide compounds has been extensively 
studied by experiments~\cite{Tak08, Lu08,Zoc08,Tor08,Par08,Ali08,Kreyssig08,Oka08}. 
For fluorine doped LaOFeAs, an increase of the superconducting $T_c$ under
pressure with a maximum value of 43~K around 4 GPa was reported~\cite{Tak08}.
For undoped $A$Fe$_2$As$_2$ compounds ($A$=Ca, Sr, Ba), which order
anti-ferromagnetically at ambient pressure, pressure induced superconductivity
up to $T_c\approx$ 29 K for $A$=Sr, Ba~\cite{Ali08} and up to $T_c\approx$ 12 K for 
$A$=Ca~\cite{Tor08,Par08} was observed.
Very recently, pressure induced superconductivity was also reported for undoped 
LaOFeAs with a maximum value of $T_c\approx$ 21 K around 12 GPa~\cite{Oka08}.

%%%%%%%%%%%%%%%%%%%%%%%% Computational details %%%%%%%%%%%%%%%%%%%%%%%%%%%

\section{Computational details}

We performed electronic structure calculations in the framework of DFT
using two high precision all electron codes, the full potential local orbital (FPLO) method~\cite{FPLO} 
and the FLAPW method implemented in WIEN2k~\cite{WIEN}. 
To ensure that our conclusions do not depend on the choice
of functional approximation to DFT, we employed both the local spin density
approximation (LSDA) and the generalized gradient approximation (GGA).
Calculations were done for different volumes in the tetragonal crystal structure (P4/nmm), as well as in the orthorhombic crystal structure (Cmma), which are the crystal structures above and below $T_s$, respectively.

We considered three different types of 
AF spin arrangements (Fig.~\ref{FIG:af123}): 
The first cell (AF1) corresponds 
to a checkerboard arrangement in the original unit cell, where nearest neighbor Fe 
atoms are aligned anti-ferromagnetically in the $xy$-plane with a 
ferromagnetic (FM) stacking along the $c$-axis. 
Second, we considered a stripe-like spin arrangement in the plane with FM
stacking along the $c$-axis (AF2) in  a $\sqrt{2}\times\sqrt{2}\times 1$
supercell. The third spin arrangement (AF3) has the
same stripe-like stacking in the plane as AF2, but in addition the spins are also arranged
anti-ferromagnetically along the $c$-axis in a $\sqrt{2}\times\sqrt{2}\times 2$
supercell. 
The experimentally observed spin arrangement~\cite{Cru08} is AF3
and has so far not been addressed by electronic structure calculations.

For all calculations, the scalar relativistic approximation 
was used.
The FPLO calculations (FPLO version 7.00-28) 
were performed in the local spin density
approximation (LSDA) in the parameterization of Perdew and Wang~\cite{Per92}.
For the {\bf k}-space integrations 512 {\bf k}-points 
in the full Brillouin zone (FBZ) were used for the structure optimization, and the convergence of the
magnetic moments and Fermi surface properties was checked with up to
32768 {\bf k}-points in the FBZ.
In the FLAPW calculations~\cite{WIEN} the exchange-correlation functional is evaluated
within the generalized gradient approximation (GGA), using the Perdew-Burke-Ernzerhof
parameterization~\cite{PBE96}. 
The muffin-tin-radii for La, Fe, As and O were chosen as 2.3, 2.15, 2.10  and 1.75 Bohr radii,
respectively.
Self-consistent calculations employed a grid of 4000 (AF1) and 2000 (AF2 and AF3) {\bf k}-points 
in the FBZ. $R_{MT} \times k_{\rm max} = 7$ was used 
as plane wave cut-off. %for the number of plane waves.

%%%%%%%%%%%%%%%%%%%%%%%% RESULTS %%%%%%%%%%%%%%%%%%%%%%%%%%%
%

\section{Results}
\subsection{Fe moment as a function of volume}

\begin{figure}
\includegraphics[width=0.48\textwidth]{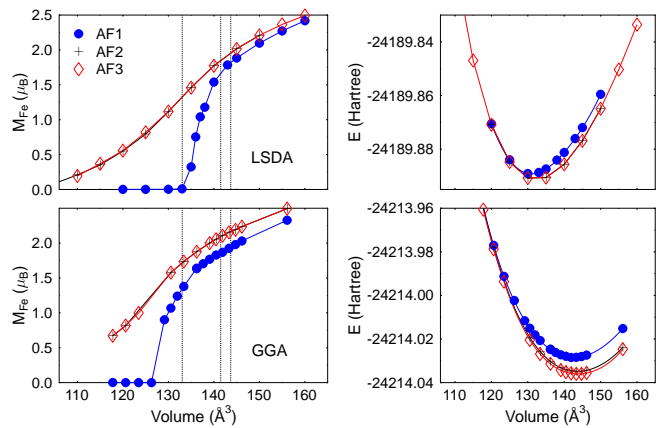}\\
\caption{\label{FIG:MvsV1}
(Color online)
Left: Fe moment as a function of volume (top: LSDA, bottom: GGA) with $c/a$, $z_{As}$ and $z_{La}$ fixed to their experimental ambient pressure values for different types of AF spin arrangement (see text). Dashed vertical lines denote the calculated LSDA equilibrium volume, the experimental volume and the calculated GGA equilibrium volume (from left to right). Right: Corresponding total energies.
}
\end{figure}
As a first step, we consider the magnetic properties of LaOFeAs as a function of volume in the tetragonal structure, with the free parameters $c/a$, $z_{\rm As}$ and $z_{\rm La}$ fixed initially to their experimental ambient pressure values~\cite{Cru08}. This allows us to distinguish between the effects of different spin arrangements, structural parameters and the influence of the exchange and correlation functional. Early electronic structure calculations (see Ref.~\onlinecite{Maz08} for an overview) showed a confusing variety of results for the magnetic properties of LaOFeAs, with values for the calculated Fe moment between almost zero and 2.6 $\mu_B$.
It soon turned out that the Fe moment is highly sensitive
to the functional 
as well as the details
of the structure and the spin arrangement
used in the calculations~\cite{Cao08,Maz08}. While calculations assuming a FM alignment
of the spins yield magnetic moments of $\approx$ 0.3 $\mu_B$ 
(almost in accidental coincidence with experiment), calculations
using the correct AF spin arrangement obtain Fe moments substantially larger than the measured
$\approx$0.4 $\mu_B$.

The variation of the Fe moment and the total energy as a function of volume for the three different spin arrangements is shown in Fig.~\ref{FIG:MvsV1}. The calculated Fe moment for the stripe-like spin arrangements AF2 and AF3 at the experimental lattice parameters is 1.87 $\mu_B$ within LSDA. The corresponding magnetic stabilization energy with respect to the nonmagnetic state is 3.2 mHartree/Fe, in good agreement with the results obtained by Mazin {\em et al.}~\cite{Maz08}. In agreement with experiment, the stripe-like spin arrangement is lowest in energy. Since the magnetic coupling between different Fe layers along the $c$-axis is weak, AF2 and AF3 are very close in energy and have also a similar electronic structure and magnetic moments, which justifies the use of the AF2 structure in earlier calculations. 

Although the three AF structures have similar Fe moments at the experimental lattice parameters, their behavior with respect to small changes of the volume is remarkably different: While the Fe moment of the checkerboard arrangement AF1 sharply drops with decreasing volume and vanishes already close to the calculated LSDA equilibrium volume, it decreases more smoothly for the stripe-like arrangements AF2 and AF3. The GGA calculations (Fig.~\ref{FIG:MvsV1}, bottom) show qualitatively the same behavior, which rules out that the observed behavior is due to a special property of a certain functional. Within GGA, the Fe moments and magnetic stabilization energies (2.11 $\mu_B$ and 6.6 mHartree at the experimental lattice parameters) are higher than in LSDA, as expected from the well-known tendency of GGA to overestimate magnetic interactions.

The remarkable sensitivity of the Fe magnetic moments on the type of AF spin arrangement and structural details shows already that LaOFeAs is on the verge to a magnetic instability with respect to changes of the volume. In the vicinity of a magnetic transition or instability, parameter-free LSDA or GGA calculations can not be expected to yield the exact value for the magnetic moment, but they should qualitatively reproduce the behavior as a function of an external parameter like pressure. However, the results shown so far do not yet fully explain the deviations between the calculated and measured Fe moment: While a sharp drop as in the case of AF1 could well explain those deviations, the Fe moments remain substantially too large in the vicinity of the equilibrium volume for the correct AF3 spin structure. 

\subsection{Structural optimization: Fe moment as a function of pressure}

Our next step is to consider structural optimizations of the free parameters $c/a$, $z_{\rm As}$ and $z_{\rm La}$ for different volumes.
In earlier publications it was pointed out that the magnetic moments
in LaOFeAs are highly sensitive to structural parameters, especially to the height
of the As position $z_{\rm As}$~\cite{Yin08,Maz08}. Under pressure, changes of the
structural parameters with respect to their ambient pressure values can be expected.
In the following we restrict ourselves to LSDA calculations, which are better suited to describe the magnetic behavior in the vicinity of a magnetic instability, due to the tendency of GGA to overestimate the magnetic interactions. LSDA calculations have for example successfully been used to predict a meta-magnetic transition in YCo$_5$ under pressure~\cite{Ros06}.

\begin{figure}
\includegraphics[width=0.45\textwidth]{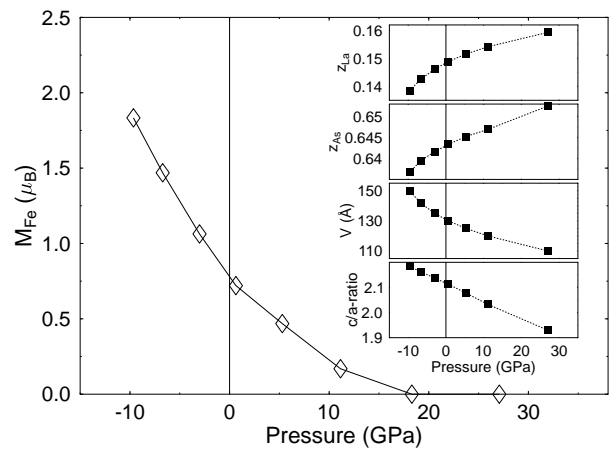}
\caption{\label{FIG:MvsV2}
Fe moment as a function of pressure for the AF3 spin-structure with optimized structural parameters. The inset shows the variation of $c/a$, $V$, $z_{\rm As}$ and $z_{\rm La}$ with pressure.
}
\end{figure}

The structural optimizations have been performed with spin-polarized calculations
in the AF3 structure. The $c/a$-ratio shrinks considerably with pressure,
reflecting a weak inter-layer coupling, and also
the internal parameters are subject to considerable changes (inset of Fig.~\ref{FIG:MvsV2}).
The Fe moment, calculated with the optimized parameters as function of pressure, is shown
in Fig~\ref{FIG:MvsV2}. Within $\pm$ 5 GPa around the calculated equilibrium volume (zero
pressure), the Fe moment drops by nearly a factor of 3. 
The calculated moment at zero pressure is 0.75~$\mu_B$ and thus still about two times larger than the experimental value. At a pressure of about 5 GPa, the calculated Fe moment coincides with the one observed in experiments. The magnetic stabilization energy decreases from 2.6 mHartree/Fe at -10 GPa to about 0.3 mHartree/Fe at ambient pressure. Spin fluctuations, which are only incompletely included in LSDA or GGA calculations, are expected to lead to a substantial suppression of the magnetic moment when the magnetic stabilization energy is of the order of 0.5 mHartree per atom~\cite{Maz08}. Hence, an even sharper reduction of the Fe moment with pressure than the one shown in Fig.~\ref{FIG:MvsV2} might be observed in experiment, although the precise effect of the spin fluctuations can not be estimated.

Up to now, we did not consider the effect of the orthorhombic lattice distortion observed at low temperatures. We find indeed a minimum in the total energy with a small deviation in the $b/a$-ratio of about 1\% from a tetragonal lattice, in agreement with experiment and an earlier report by Yildirim~\cite{Yil08}.
However, in contrast to Yildirim's work we
find only a minor influence of this orthorhombic distortion on the Fe moments $(\approx 0.05 \mu_B)$,
leaving the data shown in Fig.~\ref{FIG:MvsV2} basically unchanged.

%%%%%%%%%%%%%%%%%%%%%%%% DISCUSSION %%%%%%%%%%%%%%%%%%%%%%%%%%%

\section{Discussion}

The result of our study shown in Figs.~\ref{FIG:MvsV1} and \ref{FIG:MvsV2} predicts that LaOFeAs is
close to a magnetic instability. At a pressure of about 5 GPa, the Fe moment calculated within LSDA
coincides with the one observed in experiments. This means that LaOFeAs is on the
right side of the transition shown in Fig.~\ref{FIG:MvsV2}.
A strong increase of the Fe moment would be expected for negative pressure
conditions, which are of course not straightforward to realize in experiments. However,
it is well known that hydrogenation can lead to a sizeable increase of the volume and could
thus serve as a medium to simulate negative pressure for LaOFeAs.
Since the electronic structure of different iron arsenide compounds is quite similar~\cite{Kre08}, 
we would also expect that the behavior we found in LaOFeAs can be observed in other iron arsenide compounds. 

While finalizing this work, further studies~\cite{Kum08,Yil08a,Xie08} on the magnetic properties 
of iron arsenide compounds under pressure were reported, which support our interpretation.
Kumar {\em et al.}~\cite{Kum08} have performed a combined theoretical and experimental investigation
for SrFe$_2$As$_2$. In their LSDA calculations, they find a suppression of the magnetism at 
a critical pressure of about 10 GPa, which is slightly higher than the critical pressure extrapolated 
from experiments of 4-5 GPa. 
Yildirim~\cite{Yil08a} has performed GGA pseudopotential calculations for CaFe$_2$As$_2$ 
and LaOFeAs under pressure. His calculations for CaFe$_2$As$_2$ show a suppression of the 
Fe magnetic moment around 10 GPa (note however that CaFe$_2$As$_2$ is special due to 
the presence of a collapsed tetragonal phase). In the case of LaOFeAs, the Fe moment calculated within the GGA
pseudopotential approach is higher at ambient pressure than in our LSDA calculations and remains close to 
2 $\mu_B$ up to 10 GPa, but drops to zero at 20 GPa. However, the data shown do not allow to judge if the 
transition is smooth as in our calculations or if a sudden collapse of the moments under pressure occurs.
Finally, Xie {\em et al.}~\cite{Xie08} have performed calculations for BaFe$_2$As$_2$ under pressure
and found a suppression of the Fe moment around 13 GPa. 

An explanation of the magnetism in LaOFeAs in terms of localized magnetic moments is difficult,
if not impossible. First, the magnetic moments are soft and depend on the spin arrangement and structural
details, which is not compatible with a simple Heisenberg model. Second, the total band width of the Fe 3d
states amounts to about 7 eV. Near the Fermi energy, all five d-orbitals contribute to the density of states
(DOS), with little
admixture of As 4p states. In localized systems, crystal field splittings are a valuable tool to predict
the spin state~\cite{Zha08,Jes07}.
The related crystal field splittings of the Fe 3d states in LaOFeAs (evaluated 
from the center of gravity of the corresponding partial DOS) are well below 0.5 eV and thus much smaller
than the band width.

\begin{figure}
\includegraphics[width=0.4\textwidth]{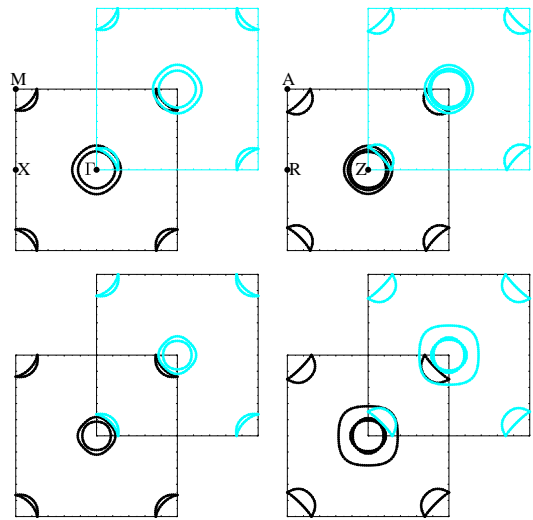}
\caption{\label{FIG:nesting}
(Color online)
Cuts through the FS of LaOFeAs for two different volumes (top: V=141.9 \AA$^3$, bottom: V=120 \AA$^3$). Cuts perpendicular to the $c$-axis in the $\Gamma$-plane (left) and in the $Z$-plane (right) are shown. To visualize nesting, the same cuts shifted by $Q=(\pi,\pi,0)$ are also drawn (light color). 
Similar nesting is observed throughout $\Gamma$-$Z$.}
\end{figure}

In an itinerant magnet like LaOFeAs~\cite{Sin08}, the magnetic state is determined by a delicate balance between
kinetic energy (favoring a nonmagnetic state) and the gain in exchange energy by spin polarization.
With increasing pressure, the bands of LaOFeAs are broadened and weight is shifted away from the
Fermi energy. A (rough) quantitative measure for this shift can be obtained from the integrated partial
DOS weighted with a Gaussian around the Fermi energy~\footnote{The width of the Gaussian has been chosen to 
be 1 eV, comparable to the size of typical exchange splittings.}, which yields a 20-30\% reduction 
from -10 to +10 GPa, where the Fe moment essentially drops from 2$\mu_B$ to almost zero.
We do not observe abrupt changes in the electronic structure in this pressure range,
consistent with the smooth (but rapid) decrease of the Fe moment in Fig.~\ref{FIG:MvsV2}.
However, close to the magnetic instability, the observed changes are sufficient to alter 
the magnetic state.

Mazin {\em et al.}~\cite{Maz08a} have pointed out that the stripe-like spin arrangement is stabilized by nesting features in the paramagnetic Fermi surface (FS). This is confirmed by our calculations, which also show that the related nesting features remain robust under pressure. The FS consists of five sheets,
with two cylindrical hole sheets around $\Gamma$ and two cylindrical
electron sheets around $M$, nested by a vector $Q=(\pi,\pi, 0)$. 
In addition, there is a hole pocket around $Z$, whose shape depends
strongly on structural details~\footnote{This hole pocket becomes a cylinder when the experimental values
for $z_{\rm As}$ and $z_{\rm La}$ are used.}. 
The larger one of the $\Gamma$-centered FS sheets becomes more three dimensional around 10 GPa, but
the topology of the Fermi surface does not change up
to pressures of at least 10 GPa, and also the nesting features are surprisingly robust (Fig.~\ref{FIG:nesting}).
The nesting is never perfect, but remains substantial throughout the considered pressure 
range, which explains the relative stability of the magnetism in the AF3 structure. %range.

\section{Conclusions}

In summary we have shown that LaOFeAs is close to a magnetic instability, which explains
the discrepancies between the values for the Fe moment found in experiment and DFT calculations.
On the basis of our calculations we expect a strong increase of the Fe moment with
increasing volume, which could  be realized for example by hydrogenation.
The Fermi surface topology and the reported nesting properties are fairly robust up to pressures
of at least 10 GPa.

\section*{Acknowledgments}

We acknowledge useful discussions with P.J. Hirschfeld and 
we thank the Deutsche Forschungsgemeinschaft for financial support
through the TRR/SFB~49 program.

%%%%%%%%%%%%%%%%%%%%%% BIBLIOGRAPHY %%%%%%%%%%%%%%%%%%%%%%%%%%%%%%%%%%%%%

%%%%%%%%%%%%%%%%%%%%%% END BIBLIOGRAPHY %%%%%%%%%%%%%%%%%%%%%%%%%%%%%%%%%
\end{document}